\documentstyle[./aabib,./psfig]{l-aa}

\thesaurus{08.02.1;  
           08.02.7;  
           08.05.3;  
           08.13.2;  
           10.07.2;  
           10.07.3}  
\newcommand{\apj}{\mbox{ApJ}}
\newcommand{\apjl}{\mbox{ApJ}}
\newcommand{\aj}{\mbox{AJ}}
\newcommand{\aap}{\mbox{A\&A}}
\newcommand{\mnras}{\mbox{MNRAS}}
\newcommand{\nat}{\mbox{Nat.}}

\newcommand{\msun}{\mbox{${\rm M}_\odot$}}

\newcommand{\rsun}{\mbox{${\rm R}_\odot$}}
\newcommand{\kms}{\mbox{${\rm km~s}^{-1}$}}

\newcommand{\paperI}{\mbox{paper~I}}

\newcommand{\nbody}{\mbox{{{\em N}-body}}}
\newcommand{\tcrss}{\mbox{${t_{\rm hm}}$}}
\newcommand{\trlx}{\mbox{${t_{\rm rlx}}$}}

\newcommand{\mavc}{\mbox{${\langle m\rangle_{\rm c}}$}}

\newcommand{\rhocore}{\mbox{${\rho_{\rm core}}$}}
\newcommand{\rcore}{\mbox{${r_{\rm core}}$}}
\newcommand{\rvir}{\mbox{${r_{\rm vir}}$}}
\newcommand{\rhm}{\mbox{${r_{\rm hm}}$}}

\def\unit#1{{\mbox{[{\rm #1}]}}}
\def\apgt{\ {\raise-.5ex\hbox{$\buildrel>\over\sim$}}\ }
\def\aplt{\ {\raise-.5ex\hbox{$\buildrel<\over\sim$}}\ }

\begin{document}


\title{Star cluster ecology III:
       Runaway collisions in young compact star clusters}

\author{Simon F.\ Portegies Zwart\inst{1}\thanks{Japan Society for the
                                                 Promotion of Science Fellow}
\and 	Junichiro Makino\inst{1}
\and    Stephen L.\ W.\ McMillan\inst{2}
\and    Piet Hut\inst{3}
}	

\offprints{Simon Portegies Zwart: spz@grape.c.u-tokyo.ac.jp}

\institute{
$^1$ Department of Information Science and Graphics, 
		College of Arts and Science, 
		University of Tokyo, 3-8-1 Komaba,
		Meguro-ku, Tokyo 153, Japan \\
$^2$ Department of Physics and Atmospheric Science, 
		Drexel University, 
                Philadelphia, PA 19104, USA \\
$^3$ Institute for Advanced Study,
                Princeton, NJ 08540, USA 
}

\date{received; accepted:}
\maketitle
\markboth{Portegies Zwart et al.\, :Star cluster ecology III}{}

\begin{abstract}
The evolution of young compact star clusters is studied using $N$-body
simulations in which both stellar evolution and physical collisions
between stars are taken into account. The initial conditions are
chosen to represent R\,136, a compact star cluster in the 30\,Doradus
region of the Large Magellanic Cloud.  The present runs do not include
the effects of primordial binaries.

We find that physical collisions between stars in these models are
frequent, and that the evolution of the most massive stars and the
dynamical evolution of the cluster are closely coupled.  In all
cases, a single star grows steadily in mass through mergers with other
stars, forming a very massive ($\apgt100\msun$) star in less than
3--4 Myr.  The growth rate of this runaway merger is much larger than
estimates based on simple cross-section arguments, mainly because the
star is typically found in the core and tends to form binaries with
other massive stars there.  The runaway is ``rejuvenated'' by each new
collision, and its lifetime is extended considerably as a consequence.
Observationally, such a star will appear in the Hertzsprung-Russell
diagram as a blue straggler.  When the runaway forms a black hole, the
binary in which it is found is usually dissociated.

We further investigate the sensitivity of the runaway to different
formulations of mass loss from high-mass main sequence stars.  We find
that, while the runaway process is less pronounced in the presence of
strong stellar winds, the basic effect persists even in the face of
large mass loss.


\end{abstract}

\section{Introduction}
Physical collisions between stars are not rare in the central regions
of star clusters or galaxies. In some star clusters, and in the cores
of many galaxies, stellar collisions are likely to play an important
role in the formation of exotic objects such as blue stragglers
(Sanders 1970; McNamara \& Sanders
1976)\nocite{san70}\nocite{1976A&A....52...53M}, X-ray binaries
(Fabian et al.~1975)\nocite{fpr75} and millisecond pulsars (Lyne et
al.\ 1987;
1988).\nocite{1987Natur.328..399L}\nocite{1988Natur.332...45L}
Collisions may also be responsible for the color gradient observed in
several post-collapse globular clusters (Djorgovski et
al.~1991)\nocite{1991ApJ...372L..41D}, and the formation of central
black holes in galactic nuclei (Quinlan \& Shapiro 1990; Quinlan et
al.~1995, Lee
1995).\nocite{1995ApJ...440..554Q}\nocite{1995MNRAS.272..605L}
Collisions in proto-clusters may be responsible for the formation of
massive stars (Bonnell et al.\ 1998)\nocite{astro-ph/9802332}, and
possibly even the entire mass spectrum (Silk \& Takahashi 1979; Allen
\& Bastien 1995; Price \& Podsiadlovski 1995)
\nocite{1979ApJ...229..242S}\nocite{1995ApJ...452..652A}
\nocite{1995MNRAS.273.1041P}

In order to quantify the effect of collisions on the evolution of
stars and the corresponding changes in the stellar population,
Portegies Zwart et al.\ (1997, hereafter \paperI)\nocite{pzhv97}
performed population synthesis calculations which included stellar
collisions.  In those calculations the stellar number density was held
constant, thus excluding the possibility of any interplay between the
dynamical evolution of the cluster and collisions between stars.

In reality, both the stars and the parent cluster evolve on comparable
time scales, and cluster dynamics and stellar evolution are quite
closely coupled.  For example, massive stars tend to segregate to the
core due to dynamical friction, increasing their collision
probability.  Their collision products are even more massive, leading
to the possibility of runaway merging (Lee 1987; Quinlan \& Shapiro
1990), \nocite{1987ApJ...319..801L} \nocite{1990ApJ...356..483Q} if
the collision rates can remain high enough in the few Myr before the
stars explode as supernovae.  However, these rates are determined by
the dynamical state of the cluster, which is in turn strongly
influenced by stellar mass loss.  The only way
to treat this intimate coupling between
stellar collisions and cluster dynamics is to perform \nbody\
simulations in which the stars are allowed to evolve and collide with
one another in a fully self-consistent way.

In this paper we report the results of a series of \nbody\ simulations
modeling young and compact star clusters, such as R\,136 in the
30\,Doradus region in the Large Magellanic Cloud.  This cluster is
particularly interesting because a strong coupling between stellar
evolution and stellar dynamics may exist. In addition to this,
excellent observational data is available.  Many
unusually bright and massive stars (e.g.\ Massey \& Hunter
1998)\nocite{MH98} are 
present in R\,136 which, due to the high central density of $10^6$ to
$10^7$ stars pc$^{-3}$, are likely to interact strongly with
each other.

The numerical method is discussed in Sect.\,2.  Sect.\,3 describes
in more detail the initial conditions for our models.  In Sect.\,4
the results are presented; they are discussed in Sect.\,5.
Briefly, we find that runaway collisions of
massive stars can occur.  The most massive star grows in mass through
merging with other stars until it collapses to a black hole.  The
growth rate of this star is much larger than estimates based on simple
cross-section arguments, because the star is typically found in the
cluster core, and tends to form binaries with other massive stars.

\section{Numerical method}
The \nbody\ integration algorithm, used in this paper, is described in
Sect.\,\ref{Sect:integrator}. In Sect.\,\ref{Sect:SeBa} we describe how the
evolution of stars are calculated; the effect of collisions on the
evolution of stars is described in Sect.\,\ref{Sect:collisions}.

\subsection{The \nbody\ integrator}\label{Sect:integrator}
The \nbody\ portion of the simulations is carried out using the {\tt
kira} integrator, operating within the Starlab software environment
(McMillan \& Hut 1996; Portegies Zwart et al.\ 1998).\nocite{mh96}
Time integration of stellar orbits is accomplished using a
fourth-order Hermite scheme (Makino \& Aarseth 1992).\nocite{ma92}
{\tt Kira} also incorporates block timesteps (McMillan 1986a; 1986b;
Makino 1991)\nocite{1986ApJ...307..126M}\nocite{1986ApJ...306..552M}
\nocite{1991ApJ...369..200M} special treatment of close two-body and
multiple encounters of arbitrary complexity, and a robust treatment of
stellar and binary evolution and stellar collisions (see below).  The
special-purpose GRAPE-4 (Makino et al.\
1997)\nocite{1997ApJ...480..432M} system is used to accelerate the
computation of gravitational forces between stars.  The treatment of
stellar mass loss is as described in Portegies Zwart et al.\
(1998).\nocite{1998A&A...337..363P} A more complete description the
Starlab environment is in preparation.

\subsection{Stellar evolution}\label{Sect:SeBa}
The evolution of stars is taken from the prescription by Portegies
Zwart \& Verbunt (1996, Sect.\, 2.1).\nocite{pzv96} However, some
changes are made to the mass loss in the main-sequence stage for
massive stars.

\subsubsection{Mass loss from main-sequence stars}\label{Sect:massloss}
The original equations from Eggleton et al.\ (1989),\nocite{eft89} on
which the stellar evolution model is based, ignore mass loss during
the main sequence stage.  However, for stars more massive than
25\,\msun, mass loss on the main sequence can be substantial.  We use
three different prescriptions to investigate the effect of
main-sequence mass loss.

The first prescription simply follows Eggleton et al., and no mass is
lost on the main-sequence.  In these cases, a massive star loses its
entire hydrogen envelope when it leaves the main-sequence and becomes
a Wolf-Rayet star.  We refer to this prescription as {\em no mass
loss}.

In the second prescription the mass loss rate for a massive
main-sequence star is taken to be constant in time, in such a way
that, as it leaves the main sequence, the star has lost its entire
hydrogen envelope.  We refer to this type of mass loss as {\em
constant mass loss}.

In the third and most realistic treatment, a massive star loses its
hydrogen envelope during the main-sequence phase according to the law
\begin{equation}
\dot{m} \propto - t^{1.5}. 
\end{equation}
This treatment is supported by model computations for massive stars by
Schaerer et al.\ (1999).\nocite{1999A&A...341..399S} We refer to this
type of mass loss as {\em moderate mass loss}.

In several cases the mass of collision products exceeds
$120$\,\msun\ (the most massive evolutionary tracks available,
see Schaerer et al.\ 1992).\nocite{1993A&AS...98..523S} Very little is
known about the evolution of such massive stars (see Stothers et al.\
1997; de Koter et al.\ 1998 and Figer et al. 1998).\nocite{1997ApJ...489..319S}
\nocite{dekoter98}\nocite{figer98} We assume that for such high
mass stars the lifetime and the radius depends weakly on mass (see
Langer et al.\ 1994).\nocite{1994A&A...290..819L} In our model a
100\,\msun\ star has a main-sequence lifetime of 3.08\,Myr; a
150\,\msun\ star lives for 2.98\,Myr.

\subsubsection{Supernovae and velocity kicks}
A star with a mass larger than 40\,\msun\ leaves a black hole after
ejecting its envelope during the main-sequence and Wolf-Rayet phases.
The mass of the black hole is computed as $m_{\rm bh} = 0.35 m_0 - 12$
\,\msun, where $m_0$ is the initial mass of the star.
For a star whose mass increases due to
collisions, $m_0$ is the highest mass reached by the star.

Stars with masses between 8\,\msun\ and 40\,\msun\ become neutron
stars.  At birth a neutron star receives a high velocity `kick' in a
random direction.  The magnitude of the velocity kick is chosen
randomly from the distribution proposed by Hartman
(1997).\nocite{1997A&A...322..127H} This distribution is flat at
velocities below 250\,\kms, but has a tail extending to several
thousand \kms.

Stars with masses less than 8\,\msun\ become white dwarfs. The mass of
the white dwarf equals the core mass of its progenitor at the tip of
the asymptotic giant branch.

\subsection{Stellar collisions}\label{Sect:collisions}
A collision is assumed to occur when two stars ($i$ and $j$) approach
each other within a distance $d = 2 (r_i + r_j)$, where $r_i$ and
$r_j$ are the radii of the stars involved.

\subsubsection{The collision product} \label{Sect:rejuvenation}
A detailed description of the treatment for collisions is given in
\paperI\ (section 3.3).  Here we summarize the prescription for
collisions between main-sequence stars.

A collision between two main-sequence stars with masses $m_i$ and
$m_j$ results in a single rejuvenated main-sequence star with mass
$m_i + m_j$.  Smooth-particle hydrodynamic simulations of collisions
between main-sequence stars indicate that at maximum a few percent of
the total mass is lost (see e.g.: Lai et al.\ 1993; Lombardi et al.\
1995; 1996).  Consequently, mass loss during the merger event is
ignored.
\nocite{1993ApJ...412..593L}\nocite{1996ApJ...468..797L}\nocite{1995ApJ...445L.117L}

The collision results in a reduction of the age of the collision
product. The age reduction factor $f_{\rm red}$ is computed from the
mass of the most massive of the two colliding stars $m_i$ and the mass
of the collision product (Meurs \& van den Heuvel 1989)\nocite{mvdh89}
\begin{equation} 
	f_{\rm red}(m_i, m_j) = \frac{m_i}{m_i + m_j} \;
			         \frac{\tau_{\rm ms}(m_i + m_j)}
                                      {\tau_{\rm ms}(m_i)}. 
\label{Eq:rejuvenation}\end{equation} 
Here $\tau_{\rm ms}(m_i)$ is the main-sequence lifetime of a star with
mass $m_i$. The new age of the collision product is computed with
$t_\star(m_i + m_j) = f_{\rm red}(m_i, m_j) t_\star(m_i)$.

As an example, suppose that, at $t=5$\, Myr, a star with
$m_i=20$\,\msun\ collides with a second star with $m_j=8$\,\msun.
Both stars lie on the main sequence, and both are experiencing
a collision for the first time (i.e.~$t_\star = 5$\,Myr for both
stars). For a 20\,\msun\ star, $\tau_{\rm ms}(20\,\msun) \sim
7.7$\,Myr. The collision results in a main-sequence star with a mass
of 28\,\msun, for which $\tau_{\rm ms}(28\,\msun) \sim 5.4$\,Myr.  The
new age of the collision product is computed using
Eq.\,(\ref{Eq:rejuvenation}), which in this case gives
$t_\star(28\,\msun) \sim 2.5$\,Myr.

\section{Selection of initial conditions}\label{Sect:Init}
We selected the initial parameters for the models to mimic class of
star clusters similar to the Galactic cluster NGC 3606 or the young
globular cluster NGC 2070 (R136) in the 30 Doradus region of the
Large Magellanic Cloud. 

\subsection{The star cluster R\,136 in the 30\,Doradus region}
The half-mass radius ($r_{\rm hm}$) of R\,136 is about 1 parsec
(Brandl et al.\ 1996),\nocite{BSB+96} and the core radius is $\sim
0.02$\,pc (Hunter et al.\ 1995).\nocite{1995ApJ...444..758H} The total
mass $M \sim 2\,\times\, 10^4\,$\,\msun.  With an assumed mean mass of
0.6\,\msun\, the cluster thus contains about 35\,000\, stars.  The
corresponding central density is of the order of
$10^6$\,\msun\,pc$^{-3}$.  The age of R\,136 is $\sim 3$--4\,Myr
(Campbell et al.\ 1992).\nocite{CHH+92} The Galactic star cluster NGC 3606 is
somewhat smaller in size and its total mass is larger resulting in a
denser core (Moffat et al.\ 1994; Drissen et
al.\,1995).\nocite{1994ApJ...436..183M}\nocite{1995AJ....110.2235D}

For both clusters, the tidal effect of their parent galaxy (for R\,136
that is the Large Magellanic Cloud) is small.  Assuming that the mass
of the LMC is $M_{\rm LMC} \sim 10^9$\,\msun\ and the distance from
the center of the LMC is $\sim 1$\,kpc, the tidal radius,
\begin{equation}
	r_{\rm t} = \left( {M \over 3 M_{\rm LMC}}
                       \right)^{1/3} R_{\rm LMC},
\end{equation}
is $\sim 20$\,pc, much larger than the half-mass radius of either
cluster, justifying our neglect of tidal effects.

Calculations are performed using 12k and 6k stars. Therefore we need
to scale the dynamical timescale and the collision cross section to
mimic the evolution of a star cluster with larger $N$. This scaling is
discussed in the following two sections.

\subsection{Scaling the dynamical timescale}
The evolution of an isolated star cluster is driven by two-body
relaxation. Therefore, we set up the initial model so that it has the
same relaxation timescale as the real cluster.  The relaxation time is
calculated with
\begin{equation}
	t_{\rm rlx} \propto {N \over \log_{10} (\gamma N)} \tcrss.
\label{Eq:trlx}\end{equation}
Here $\gamma$ is a scaling factor, introduced to model the effects of
the cut-off in the long range Coulomb logarithm (see Giertz \& Heggie
1996; 1994).\nocite{GH96}\nocite{1994MNRAS.268..257G}
Here \tcrss\ is the half-mass crossing time of the cluster is
\begin{equation}
\tcrss \simeq 57 \left( {[\msun] \over M} \right)^{1/2}
                 \left( {\rhm \over [{\rm pc}]}\right)^{3/2} \, \unit{Myr}.
\label{Eq:thc}\end{equation}
Here $r_{\rm hm}$ is its half mass radius.  

The radius of the scaled model $r_{\rm model}$ is then computed by 
substitution of Eq.\,(\ref{Eq:trlx}) into Eq.\,(\ref{Eq:thc}):
\begin{equation}
	r_{\rm model} = \left( { N_{\rm real} \over N_{\rm model} } 
   		        \right)^{1/3}
			\left( { \ln (\gamma N_{\rm real}) \over \ln
   		        (\gamma N_{\rm model}) } 
			\right)^{-2/3}
			r_{\rm real}.
\label{Eq:rmodel}\end{equation}
Here $r_{\rm real}$ is the radius of the real cluster.  We decided to
use $\gamma =1$.  Note that Eq.\,(\ref{Eq:rmodel}) is relatively
insensitive to the exact choise of $\gamma$.

\subsection{Scaling the collision cross section}\label{Sect:collscale}
The model clusters should have the same collision rate per star as the
real cluster.  Scaling the initial conditions to assure that the model
cluster has the same relaxation time causes it to be larger than the
real system (Eq.\,\ref{Eq:rmodel}).  The correct collision rate per
star is then obtained by scaling the sizes of the stars themselves.

The number of collisions per star per unit time is given by
\begin{equation}
 	n_{\rm coll} \propto n_c \sigma v.
\end{equation}
Here $n_c$ is the number density of the stars in the core, $\sigma$ is
the collision cross section (for approach within some distance $d$),
and $v$ is the velocity dispersion.  These are given by the following
proportionalities:
\begin{eqnarray}
	n_c 	&\propto& 	{N \over r_c^3}, \nonumber	\\
	\sigma 	&\propto& 	d^2 + {d \over v^2}. 
\end{eqnarray}
Here $r_c$ is the cluster's core radius.  
We will neglect the $d^2$ term in the cross section.
Expressed in real units and assuming scaling according to
Eq.\,(\ref{Eq:rmodel}), we may write
\begin{eqnarray}
	{N \over r_c^3} &\propto& N^2, \nonumber	\\
	v 	        &\propto& 	N^{2/3}.
\end{eqnarray}
The number of collisions then becomes
\begin{equation}
 	n_{\rm coll} = d N^{4/3}. 
\end{equation}

The distance at which a collision occurs therefore scales as
\begin{equation}
 	d \propto N^{-4/3}. 
\end{equation}

\subsection{The models}
\begin{figure}
\hspace*{1.cm}
\psfig{figure=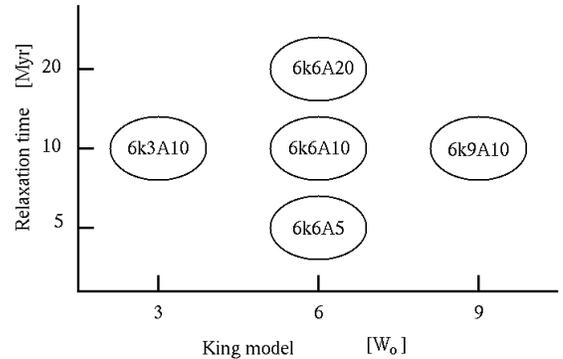,width=7.5cm,angle=0}
\caption[]{Initial parameters for the 6k runs.
}
\label{fig:models}
\end{figure}

We performed 4 runs with 12k stars and 7 runs with 6k stars.  All 12k
runs start from the same initial conditions, except for the treatment
of mass loss on the main sequence.  For 6k models we also change
the initial relaxation time and the initial density distribution.
Table\,\ref{Tab:init} and Fig.\,\ref{fig:models} summarize the initial
conditions.

All simulations start at $t=0$ by assigning masses of stars between
0.1\,\msun\ and 100\,\msun\ from the mass function suggested for the
Solar neighborhood by Scalo (1986).\nocite{sca86} At the high-mass
end this mass function is rather steep;
\begin{equation}
	N(m) \propto  m^{-2.82},
\end{equation}
and the mass function turns over at around 0.3\,\msun.  The median
mass of this mass function is about 0.3\,\msun, and the mean mass
is about 0.6\,\msun.  We generate the mass distribution using the
random sampling technique suggested by de la Fuente Marcos et al.\
1997)\nocite{EFMAK97}.

The initial density profile and velocity dispersion for the models
with 12k stars are taken from a King (1966) model with $W_0 =
6$.\nocite{kin66} We chose $\rvir = 0.31$\,pc, which results in a core
radius $\rcore \approx 0.072$\,pc and a core density of $\rhocore \sim
3.6\,\times 10^{5}\,\msun\,{\rm pc}^{-3}$. The central velocity
dispersion for these models is about 8.7\,\kms\ and the initial
half-mass relaxation time $\trlx \sim 10$\,Myr.

The names used to identify our models are defined as follows: We start
with the number of stars: 12k for models with 12288 stars, and 6k for
models with 6144 stars.  The next integer identifies the selected
value for $W_0$.  The next letter indicates the stellar mass loss
model: `A' for no mass loss, `B' for constant mass loss and `C' for
moderate mass loss (see Sect.\,\ref{Sect:massloss}).  The final number
gives the initial half-mass relaxation time in millions of years.  Two
12k6C10 models were computed.  To distinguish between them, the second
is identified as 12k6C10$^\prime$.  All computations were continued
until $t=10$\,Myr.

\begin{table*}[ht]
\caption[]{ Overview of initial conditions for the runs with $N=12$\,k
(first four rows) and $N=6$\,k (last 7 rows).  The first column gives
the model name, starting with the number of stars: `12k' for the 12k
models and `6k' for the 6k models.  The following columns give the
number of stars, the initial dimensionless King $W_0$, the relaxation
time \trlx\ and the crossing time at the half-mass radius $t_{\rm
hm}$\, (both in millions of years). The next columns give the initial
core radius \rcore, the half-mass radius \rhm\ (both in parsec) and
the average core density (in \msun/pc$^3$). The last three columns
give the number of stars initially more massive than 19\,\msun\ (the
turn-off mass for a 10\,Myr old star cluster) and 25\,\msun (the mass
limit for a Wolf-Rayet star), and the mass of the initially most massive star
(in \msun) in the cluster.  }
\begin{flushleft}
\begin{tabular}{lrcr|crlrccr} \hline
Model &$N$&$W_0$&\trlx&$t_{\rm hm}$&\rcore&\rhm
			&$\log$\,\rhocore& \multicolumn{2}{c}{$N(M>x)$} 
		& $M_{\rm max}$   \\ 
      &   &         &[Myr] & [Myr]&
		\multicolumn{2}{c}{[pc]}&
			[$\msun/{\rm pc}^3$] & 
			19\,\msun & 25\,\msun 
		& [\msun]   \\ \hline
12k6A10 &12288& 6& 10&0.082&0.07 &0.25 & 5.55 & 8& 5 & 57 \\
12k6B10 &12288& 6& 10&0.082&0.07 &0.25 & 5.56 & 7& 4 & 63 \\
12k6C10 &12288& 6& 10&0.085&0.07 &0.26 & 5.61 & 6& 3 & 70 \\ 
12k6C10$^\prime$
      &12288& 6& 10&0.082&0.07 &0.25 & 5.60 &12& 9 & 86 \\ \hline
6k6A5  & 6144& 6&  5&0.076&0.05 &0.19 & 5.64 & 1& 0 & 22\\
6k3A10 & 6144& 3& 10&0.156&0.15 &0.31 & 4.53 & 4& 3 & 66 \\
6k6A10 & 6144& 6& 10&0.158&0.10 &0.31 & 5.02 & 2& 0 & 21 \\
6k9A10 & 6144& 9& 10&0.156&0.02 &0.36 & 6.36 & 6& 4 & 37 \\
6k6A20 & 6144& 6& 20&0.314&0.14 &0.47 & 4.44 & 0& 0 & 15 \\ 
6k6C10 & 6144& 6& 10&0.157&0.09 &0.30 & 4.94 & 3& 2 & 49 \\
6k6C20 & 6144& 6& 20&0.314&0.14 &0.48 & 4.45 & 3& 1 & 32 \\ \hline
\end{tabular}
\end{flushleft}
\label{Tab:init} \end{table*}

\section{Results}
In this section we describe our results.  Stellar collisions and the
growth of massive stars are discussed in Sect.\,\ref{Sect:runaway};
Sect.\,\ref{Sect:evolution} discusses evolution of the cluster structure,
while Sect.\,\ref{Sect:anatomy} gives a detailed description of the
evolution of the runaway collision product in one particular model
(12k6A10).  The results of the models with 6k stars are described in
Sect.\,\ref{Sect:6kmodels}.

\begin{table}[ht]
\caption[]{Results on the $N=12$\,k runs.  The first column gives the
model name.  The following columns give the time of the first
collision, the total number of collisions in the computation, the
number of collisions in which the most massive star participates
($N_{M_{\rm max}}$) and the number of supernovae
which occurred during the computations.
}
\begin{flushleft}
\begin{tabular}{l|lcrr} \hline
Model 	&$t_{\rm 1^{st} coll}$
	&$N_{\rm coll}$
	&$N_{M_{\rm max}}$
	&$N_{\rm sn}$ \\ 
     & [Myr]  && \\ \hline
12k6A10    &1.7&11 &11&  5 \\ 
12k6B10    &0.9&10 & 5&  2 \\ 
12k6C10    &0.3&21 &15&  4 \\ 
12k6C10$^\prime$
         &1.1&15 &10&  9 \\ \hline
\end{tabular}
\end{flushleft}
\label{tab:12kmodels} \end{table}

\begin{figure}
\hspace*{1.cm}
\psfig{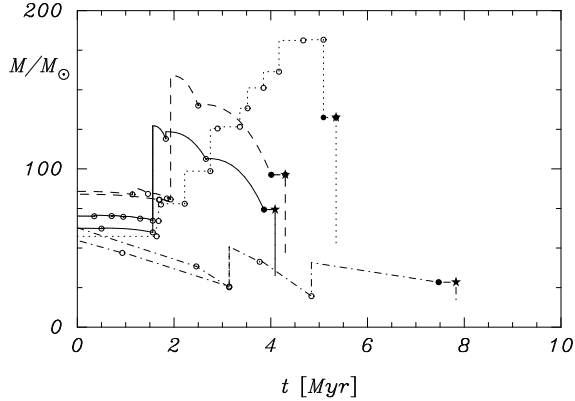}
\caption{Diagram of the encounter history for the most massive stars
in models 12k6A10 (dotted line), 12k6B10 (dash dotted line), and
models 12k6C10 (solid line) and 12k6C10$^\prime$ (dashed line).  The
horizontal axis gives time in millions of years, the vertical axis
shows the mass of the collision product. The ``$\bullet$'' symbol
indicates the moment a star leaves the main sequence and becomes a
Wolf-Rayet star. The ``$\star$'' indicates the moment the star
collapses to a black hole.  The vertical line after the $\star$ shows
the mass lost in the supernova. The evolution of the star after
the supernova is not shown.  Each collision is indicated with a small
{\small $\circ$}.  For some models
two stars are initially displayed.  In these cases, the two stars
ultimately merge, but only after each experiences collisions with less
massive stars.  For example, in model 12k6C10, the 62\,\msun\ star
experiences a collision with a lower mass star before it encounters
the 70\,\msun\ star, which itself had experienced 4 collisions prior
to the merger.}
\label{fig:N12Mcoll}
\end{figure}

\subsection{Runaway merging}\label{Sect:runaway}
Figure\,\ref{fig:N12Mcoll} gives the evolution of the mass of the most
massive star for all $N=12$\,k runs.  In all cases, more than 10
collisions occurred during the first 5\,Myr.  In model 12k6A10 (no
mass loss), a total of 11 collisions resulted in a star with a mass of
182\,\msun\ at the point when it left the main sequence.

 From Fig.\,\ref{fig:N12Mcoll} we can see that the lifetime of the
most massive star is considerably larger than its ``natural''
main-sequence lifetime of about 3\,Myr.  This extension of the
lifetime is caused by rejuvenation through merging.

In other models, two massive stars collide with relatively low-mass
stars before they eventually find each other.  In models 12k6C10 and
12k6C10$^\prime$ this happens at $t \sim 1.7$\,Myr, in model 12k6B10
at $ t \sim 3$\,Myr.

Table\,\ref{tab:12kmodels} gives information about collisions in the
$N=12$\,k models.  It typically requires about 1 million years\, --the
time needed for the most massive stars to segregate to the
core--\, before the first collision occurs (see
Table\,\ref{tab:12kmodels}).  These most massive stars participate in
more than 70\% of all collisions.  The most massive star in model
12k6C10 experiences its first collision at an earlier epoch because
that star happened to be born in the core.

The number of collisions occurring in these runs is far larger than
simple theoretical predictions.  The rate at which stars in a cluster
experiences collisions can be estimated as $n_c^2 \langle \sigma v
\rangle$ (Spitzer 1987).\nocite{spi87} Here $\sigma$ is the collision
cross section, $v$ is the velocity dispersion in the cluster and $n_c$
is the number density of stars.  Following the derivation in \paperI\
by adopting a Maxwellian velocity distribution with velocity
dispersion $\langle v \rangle$ and cross section $\sigma \propto
md/v^2$, the number of collisions is the cluster per Myr is expressed
as (see \paperI, Eq.\,14)
\begin{eqnarray}
	\Gamma &\approx 0.36& \left(\frac{\rhocore}
			       {10^4 \msun {\rm pc}^{-3}}\right)^2 
		            \left(\frac{\rcore}{{\rm pc}}\right)^3
			    \left(\frac{2 \langle m
\rangle}{\msun}\right) 
\nonumber \\
 & &		           \left(\frac{d}{\rsun}\right) 
		            \left(\frac{\kms}{\langle v \rangle}\right)
			\;\; [{\rm Myr}^{-1}].
\label{Eq:rate}\end{eqnarray}
Here $\langle m \rangle$ is the mean stellar mass.

For the $N=12$\,k runs this results in about 0.3 collision/Myr, or
about 3 collisions during the entire simulation, assuming that the
cluster parameters do not change in time. As will be discussed in
Sect.\,\ref{Sect:evolution} below, the core density in fact drops by
about an order of magnitude during the first 4\, million years.  If we
take this effect into account, the expected number of collisions is
less than unity.  The actual number of collisions in the 12k
simulations exceeds 10.  The major cause of this large discrepancy is
mass segregation, which concentrates massive stars in the core.

\begin{figure}[ht]
\hspace*{1.cm}
\psfig{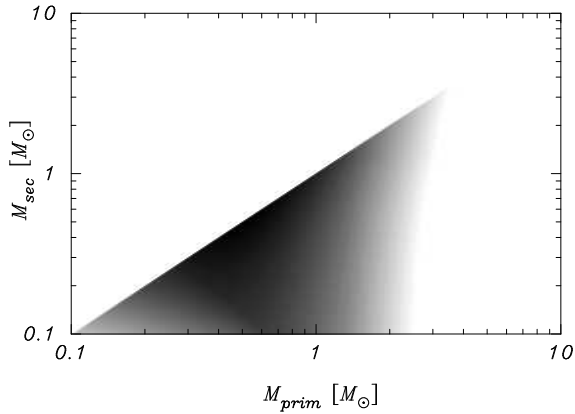}
\caption[]{Expected relative collision rate between a primary and a
secondary star for the adopted mass function and mass-radius relation
for main-sequence stars.  The velocity dispersion is taken the same
for all masses and Eq.\,(\ref{Eq:rate}) is used to compute the relative
encounter probabilities.  The shading in linear in the encounter
probability, with darker shades for higher probabilities.  Since the
probability distribution is symmetric line the axis of equal mass,
only the lower half is displayed.}
\label{fig:Renc}
\end{figure}

\begin{figure}[ht]
\hspace*{1.cm}
\psfig{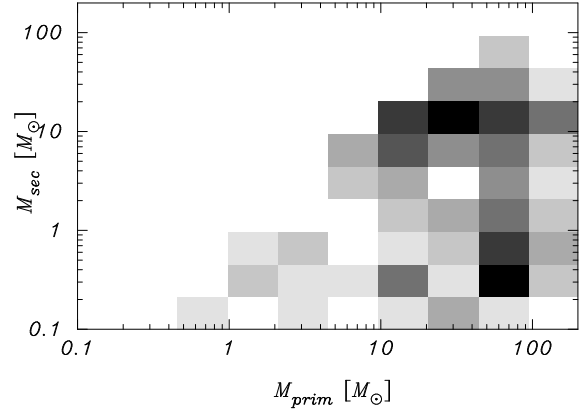}
\caption[]{ Primary and secondary masses of colliding stars for all
	simulations.  A total of 125 collisions are combined in this
	graph.  Note that the axes extend to 200\,\msun, compared with
	10\,\msun\ in Fig\,\ref{fig:Renc}.  }
\label{fig:enc12k6kN12k}
\end{figure}

The importance of mass segregation is illustrated in
Figs.\,\ref{fig:Renc} and \ref{fig:enc12k6kN12k}.
Figure\,\ref{fig:Renc} gives the theoretical probability distribution
for stars with mass $m_{\rm prim}$ to collide with lower-mass stars of
mass $m_{\rm sec}$, for the initial Scalo (1986) stellar mass distribution.
Figure\,\ref{fig:enc12k6kN12k} presents the distribution of collisions
actually observed in our simulations.

The differences between Fig.\,\ref{fig:Renc} and
Fig.\,\ref{fig:enc12k6kN12k} are striking. In the $N$-body
simulations, massive stars completely dominate the collision rate,
while theory predicts that the majority of collisions should occur
between stars of relatively low mass ($\sim 0.7$\,\msun).

\begin{figure}
\hspace*{1.cm}
\psfig{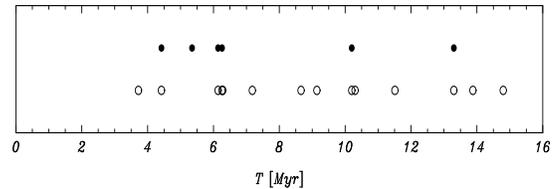}
\caption[]{Upper symbols ($\bullet$): Supernova history for model
12k6A10 over the first 16\,Myr of the evolution of the stellar system.
Lower symbols ($\circ$): the expected (scheduled) supernovae if
dynamics did not play a role.}
\label{fig:SN_N12A}
\end{figure}

Figure\,\ref{fig:SN_N12A} shows the effect of merging on the
statistics of supernovae. The total number of supernovae is reduced
due to merging of massive stars, and the explosions are delayed
relative to expectations because of rejuvenation.

\subsection{Evolution of cluster structure}\label{Sect:evolution}
\begin{figure}
\hspace*{1.cm}
\psfig{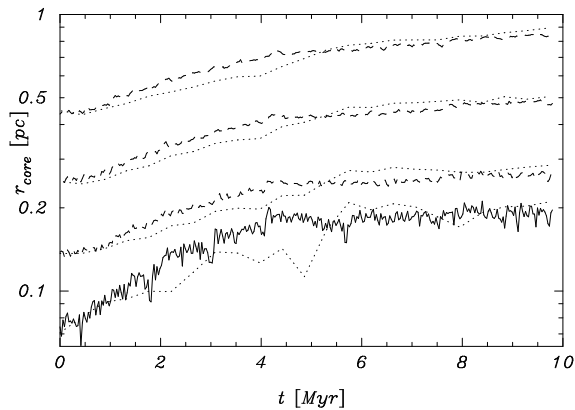}
\caption[]{Core and Lagrangian radii (in parsec) as functions of time
(in Myr) for models 12k6A10 (dotted lines) and 12k6C10 (solid and
dashed lines).  The lowest lines represent the core radius, the higher
lines give the 25\%, 50\% and 75\% Lagrangian radii.  }
\label{fig:RlagrAD}
\end{figure}

The core and the cluster as a whole expand with time (see
Fig.\,\ref{fig:RlagrAD}; for technical reasons, fewer snapshots of the
first model were stored, leading to lower temporal resolution in the
data displayed here).  For model 12k6A10 this expansion is almost
completely driven by binary heating.  For the other models, mass loss
in the stellar winds of the massive stars also drives the expansion.
Although in these models the cluster loses a modest 4\% of mass in the
first 10\,Myr, this mass is lost from deep inside the potential well
of the cluster and affects the dynamics significantly.

\begin{figure}
\hspace*{1.cm}
\psfig{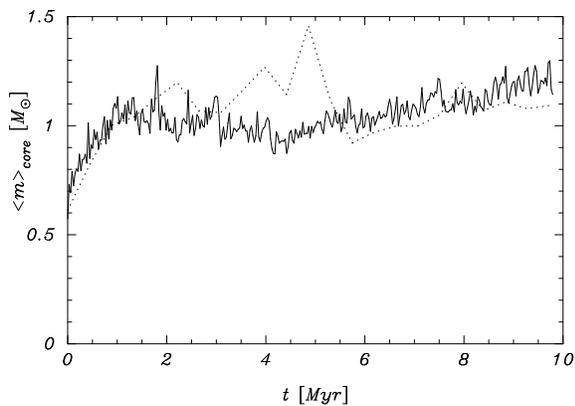}
\caption[]{Mean stellar mass in the core as a function of time for
models 12k6A10 (dots) and 12k6C10 (solid). The mean mass in the
cluster is about 0.61\,\msun, and decreases by only a few percent over
the course of the simulation.  }
\label{fig:mmcore}
\end{figure}

Figure\,\ref{fig:mmcore} shows the evolution of the mean stellar mass
in the core.  The initial mean mass in the cluster is $\langle
m\rangle \approx 0.6$\,\msun.  In the core, $\langle m \rangle$
increases to about 1.1\,\msun\ in 1\,Myr due to mass segregation.  In
model 12k6A10, $\langle m \rangle_{\rm core}$ keeps increasing until a
maximum is reached at $t \sim 5$\,Myr. In model 12k6C10, the maximum
is reached at $t\sim 2$\,Myr.  This difference is in part due to the different
treatments of mass loss from main-sequence stars.

\begin{figure}
\hspace*{1.cm}
\psfig{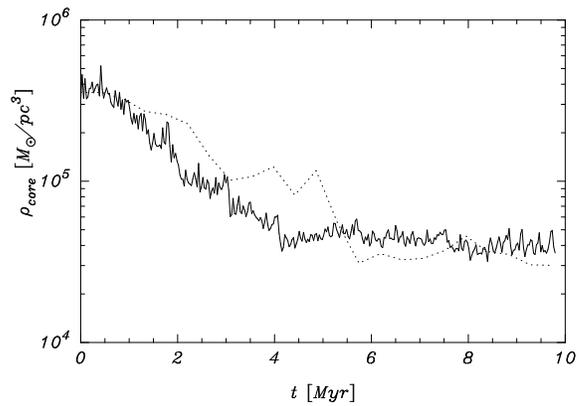}
\caption[]{Central density (in $\msun {\rm pc}^{-3}$) as a function of
time for models 12k6A10 (dotted line) and 12k6C10 (solid line).  }
\label{fig:rhocore}
\end{figure}

Figure\,\ref{fig:rhocore} shows the evolution of the core density
\rhocore\ for models 12k6A10 and 12k6C10.  The expansion of the
core causes \rhocore\ to decrease during the first 4 million years.

\subsection{Anatomy of a collision sequence}\label{Sect:anatomy}
\begin{figure*}
\hspace*{1.cm}
\psfig{figure=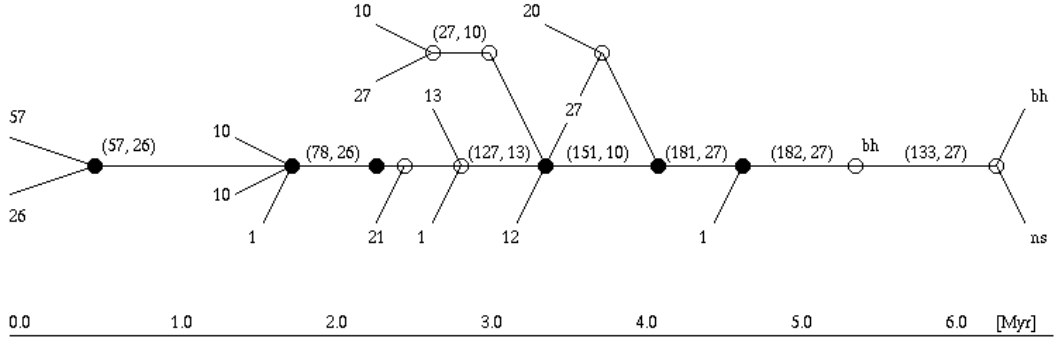,bbllx=30pt,bblly=300pt,bburx=580pt,bbury=490pt,height=5.0cm}
\caption[]{Reaction network from model 12k6A10, showing the evolution
of the most tightly bound binary, which also contains the massive
runaway merger.  Time runs from left to right.  Each line represents a
star; the numbers attached to the lines give the masses of the stars
in \msun.  When two lines join (marked by circles), a binary is
formed.  A filled circle indicates that a strong interaction with
another binary (not further specified) takes place. The masses of the
binary components, rounded to \msun, are indicated within braces.  The
difference between the number of lines entering and leaving a circle
indicates the number of collisions occurring.  The 27\,\msun\ star
returns repeatedly to encounter the binary which contains the runaway
merger. The ``{\tt bh}'' slightly after $t=5$\,Myr indicates the
collapse of the runaway merger to a black hole. The binary is finally
dissociated by a supernova which ejects a neutron star {\tt ns} (the
former 27\,\msun\ star) and a black hole (the runaway merger).  }
\label{fig:anatomy}
\end{figure*}

The first binaries in this model are formed shortly after the start of
the simulation. One of these binaries is formed from the most massive
star, of mass 57\,\msun, and a companion of 26\,\msun, with an initial
semi-major axis of 0.03\,pc. After more than ten collisions, the
57\,\msun\ star grows to 182\,\msun, before turning into a black hole
of 133\,\msun.

As an illustration of the merger process, Fig.\,\ref{fig:anatomy}
depicts a schematic reaction network of the runaway merger in model
12k6A10.

\begin{figure}
\hspace*{1.cm}
\psfig{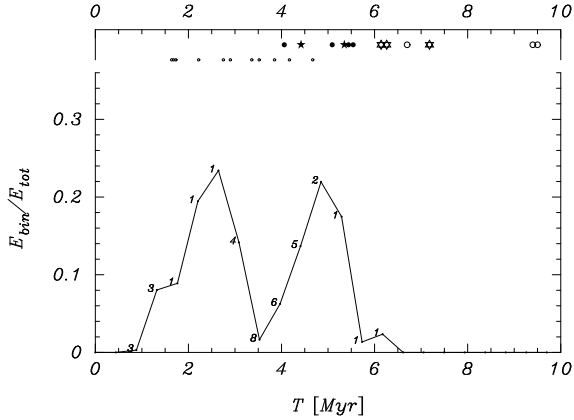}
\caption[]{Energy stored in hard ($E_b > 5kT$, with $kT$ the mean
stellar kinetic energy) binaries
as a function of time for model 12k6A10.  The numbers indicate the
number of hard binaries at that instant; they are printed only at the
moment they change.  Small open circles at the top of the figure
indicate the moments at which collisions occur.  The larger symbols
above them indicate the start of a Wolf-Rayet phase ``$\bullet$'', the
formation of a black hole ``$\star$'',and the formation of a neutron
star (open star); a ``$\circ$'' indicates when a star starts its ascent
of the super giant branch. }
\label{fig:N12A_Ebin}\end{figure}

At about 1\,Myr the binary containing the most massive star encounters
another binary. This second binary is dissociated, and its component
stars are ejected from the cluster.  The effect of this is noticeably
in Fig.\,\ref{fig:N12A_Ebin} in a steep rise in the binding energy of
the remaining binary.  Subsequently, a series of encounters with
single stars results in three collisions involving the most massive
star.  The runaway collision product, still the member of a binary,
devours two other stars between $t=2$\,Myr and 3\,Myr.

In the meantime a 27\,\msun\ star has
formed a binary with another star. 
In the core an encounter between the two binaries results in a
collision between the runaway merger and its 13\,\msun\
companion. This encounter dissipates most of the binding energy in the 
binary (see Fig.\,\ref{fig:N12A_Ebin}). The 27\,\msun\ star is ejected 
from the core to return again as the member of a binary.

The 27\,\msun\ star encounters the collision product again, the new
companion of the 27\,\msun\ star collides with the runaway merger, and
in addition the runaway merger collides again with its own companion; the
27\,\msun\ star takes its place.

Slightly after $t \simeq 5$\,Myr the runaway merger collapses to a
black hole after having consumed another 1\,\msun\ star.  The binary
survives but shortly afterward its companion (the 27\,\msun\ star)
explodes in a supernova.  The neutron star (remnant of the 27\,\msun\
star) receives a high velocity kick of $\sim 100$\,\kms, dissociating
the binary.  The black hole (remnant of the runaway merger) is also
ejected.

\begin{figure}[ht]
\hspace*{1.cm}
\psfig{figure=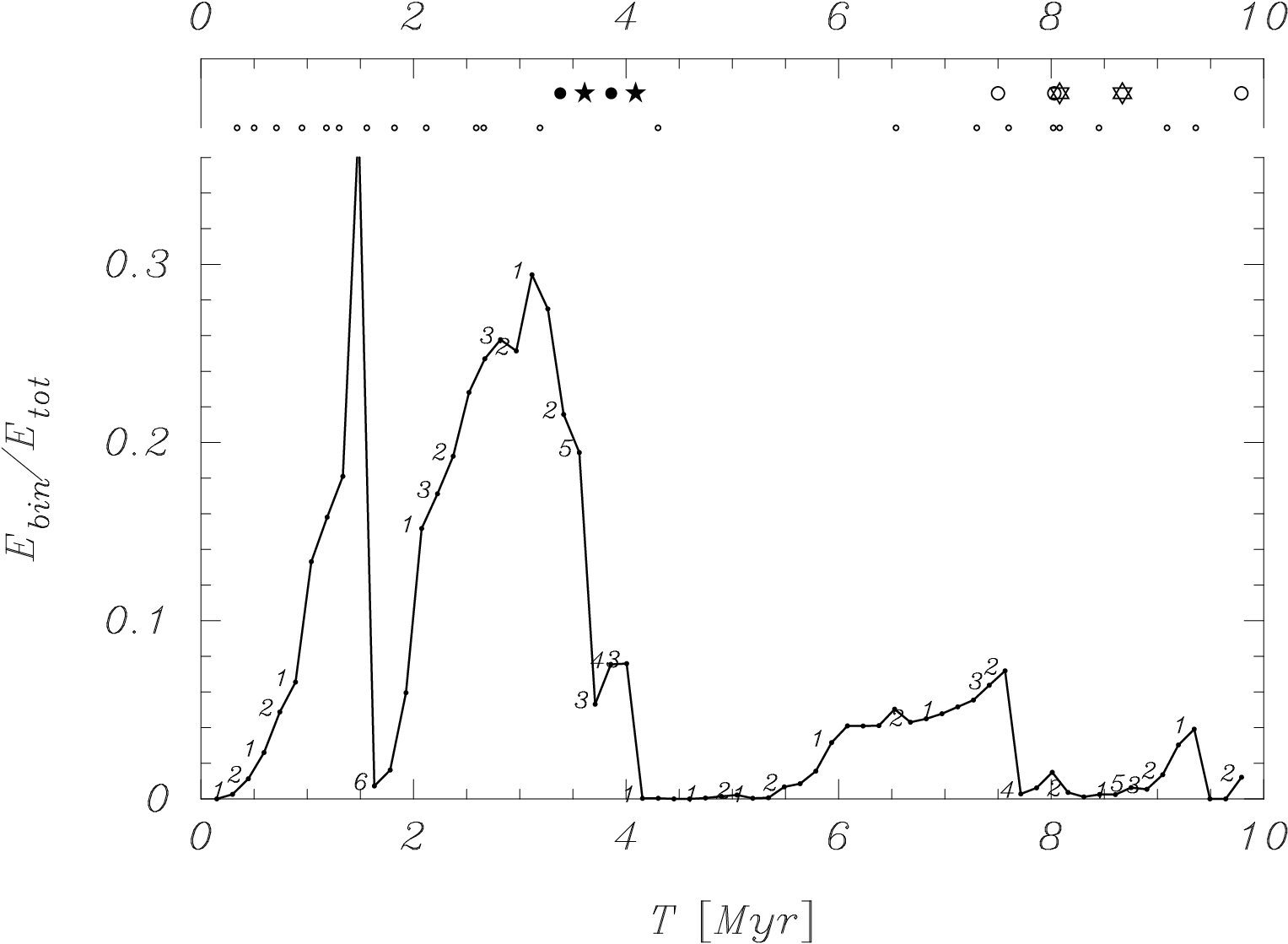,width=7.5cm,angle=-90}
\caption[]{Energy stored in hard ($E_b > 5kT$) binaries as a
function of time for model 12k6C10.  See Fig.\,\ref{fig:N12A_Ebin} for
an explanation of the symbols.}
\label{fig:N12D_Ebin}\end{figure}

\begin{figure}[ht]
\hspace*{1.cm} \psfig{figure=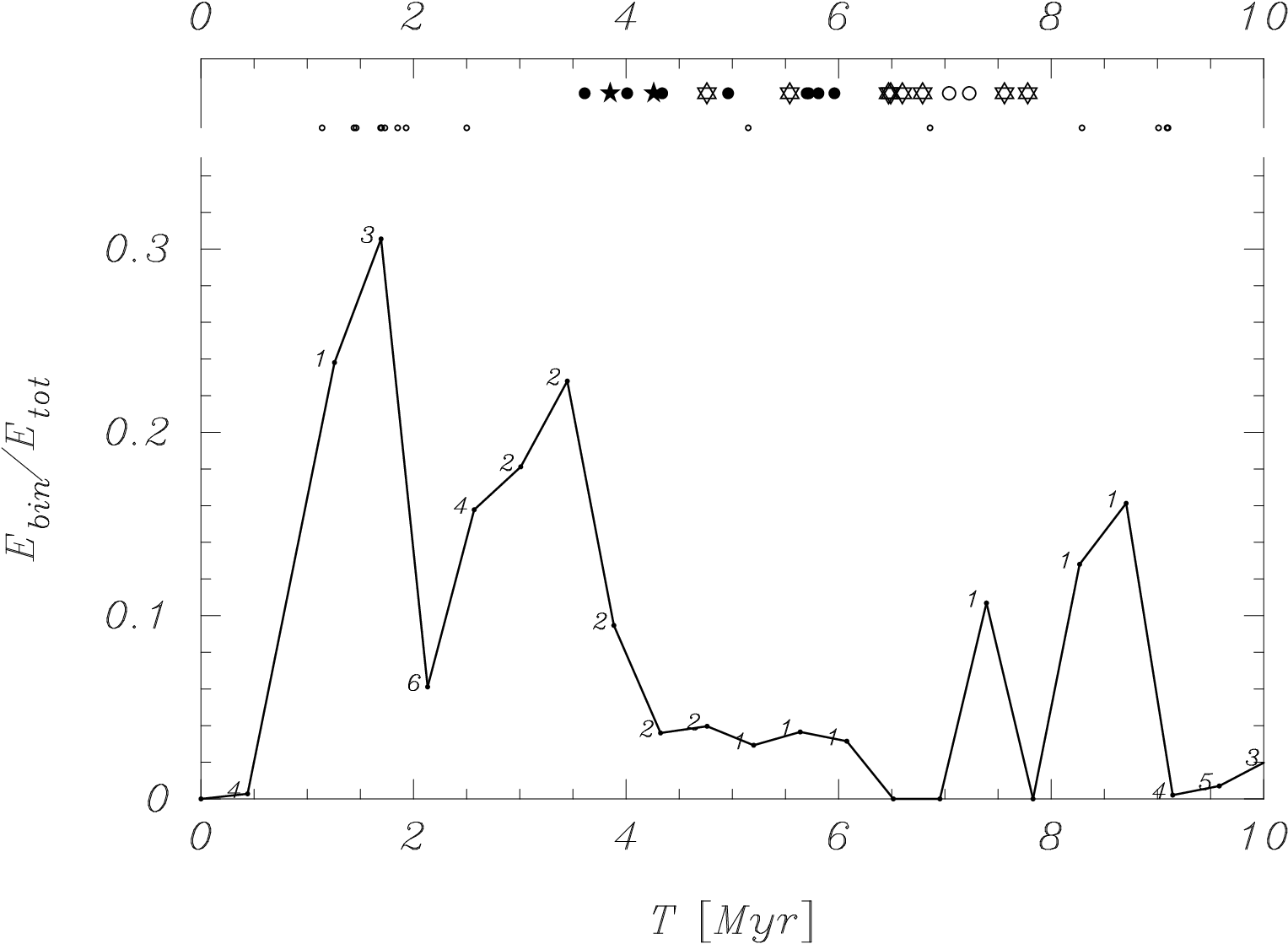,width=7.5cm,angle=-90}
\caption[]{Energy stored in hard ($E_b > 5kT$) binaries as a
function of time for model 12k6C10$^\prime$.  See
Fig.\,\ref{fig:N12A_Ebin} for an explanation of the symbols.}
\label{fig:N12C_Ebin}\end{figure}

Figure\,\ref{fig:N12A_Ebin} plots the binding energy of the binaries
and indicates some other important events for model 12k6A10.  Similar
figures are presented for model 12k6C10 in Fig.\,\ref{fig:N12D_Ebin}
and for model 12k6A10$^\prime$ in Fig.\,\ref{fig:N12C_Ebin}

Once binaries form, they gradually become harder until a maximum
binding energy is reached.  At that point the primary coalesces with
its companion, removing its binding energy from the system.  The
single star remaining after the merger captures a new companion, and
the process repeats itself.  Note that the term ``binary'' may be
somewhat misleading here, as the runaway merger is usually the primary
of a multiple system.  The 8 ``binaries'' at $t=3.5$\,Myr in model
12k6A10 are in fact a hierarchical system of 7 lower mass companions
orbiting the runaway, which at that instant has a mass of 151\,\msun.

Most collisions occur between a member of a hard binary and an
incoming star.  Following the collision, the binary becomes softer.
This is most clearly visible in Figs.\,\ref{fig:N12A_Ebin} and
\ref{fig:N12C_Ebin}.  During an episode without collisions, the
binding energy of binaries rises at a rate of $\sim 300kT$ (about
0.2\% of the binding energy of the cluster) per million years.

\subsection{Results of the 6k models}\label{Sect:6kmodels}
Table\,\ref{tab:6kmodels} provides information on the calculations
with 6k stars.  The average number of collisions for the 6k runs with
an initial relaxation time of 10\,Myr is $7.0 \pm 2.4$.  For the 12k
runs the average number of collisions is $14 \pm 4$.  The fact that
the collision rate per star per unit time is about the same in both
sets of models suggests that we may be able to extrapolate our results
to larger numbers of stars.

\begin{table}
\caption[]{Overview of the 6k runs.  The first five rows give
information on the computations performed with no mass loss; the
remaining two are from moderate mass loss runs.  The first column
gives the model name.  Subsequent columns give the time of the first
collision, the total number of collisions, and the number of
collisions in which the runaway merging star
participates ($N_{M_{\rm max}}$).  The final column gives the number
of stars experiencing a supernova during the computation.}
\begin{flushleft}
\begin{tabular}{l|lcrr} \hline
Model   &$t_{\rm 1^{st} coll}$
	&$N_{\rm coll}$
	&$N_{M_{\rm max}}$
	&$N_{\rm sn}$ \\ 
      & [Myr] \\ \hline
6k6A5  & 1.2&21&21    &1\\ 
6k3A10 & 2.6& 4&1     &3\\
6k6A10 & 2.1& 9&7     &1\\       
6k9A10 & 0.6& 6&4     &6\\ 
6k6A20 & 5.2& 3&0     &0\\ \hline 
6k6C10 & 1.2& 9&5     &1\\ 
6k6C20 & 2.7& 5&3     &1\\ \hline 
\end{tabular}
\end{flushleft}
\label{tab:6kmodels} \end{table}

The results for models with different initial relaxation times
indicate that the collision rate is indeed inversely proportional to
the relaxation time, consistent with Eq.\,(\ref{Eq:rate}).  On the other
hand, the initial central density has a rather small effect on the
total number of collisions, even though the densities range over two
orders of magnitude.

\begin{figure}[ht]
\hspace*{1.cm}
\psfig{figure=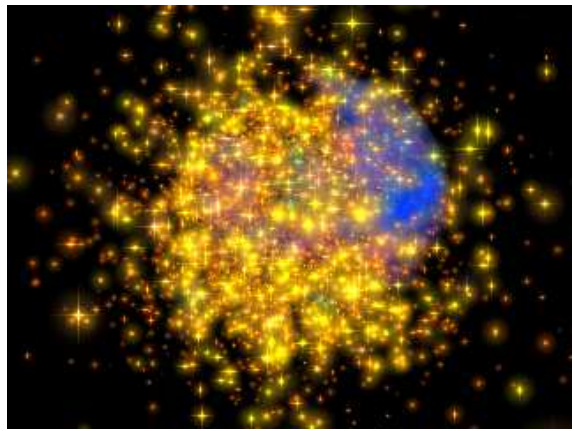,width=7.5cm}
\caption[]{ Picture of one of the simulated clusters (model 12W6A10)
from a distance of 2 parsec. An animation of the cluster can be found
at {\tt http://www.sns.ias.edu/$\sim$starlab/research/} or at the mirror
{\tt http://www.astro.uva.nl/$\sim$spz/30Doradus.html}. \\ The colors of
the stars represent temperature.  A collision product flares up in
bright white. A supernova produces a bright violet star together with
a slowly expanding shell-like structure.  }
\label{fig:animation}\end{figure}

Figure\,\ref{fig:animation} shows a picture of the simulated cluster
(model 12W6A10) from a distance of 2 parsec. An animation of the
cluster can be seen at the following address: 
{\tt http://www.sns.ias.edu/$\sim$starlab/research/30Doradus/} 
(http stands for uniform resource locator.
The mpeg (Moving Picture Expert Group) animation shows the evolution
of star cluster 12k6A10 from birth to about 7\,Myr. In the beginning we
look at the density center of the cluster from a distance of 30\,pc
($\sim 100\,r_{\rm hc}$). Then we zoom in with a velocity of about
29\,\kms\ to a distance of 3\,pc ($\sim 10\,r_{\rm hc}$).  

\section{Discussion}
We have studied collisions in young star clusters with 6144 stars and
12288 stars. The initial conditions of the models are selected to
mimic stellar systems with a larger number of stars.  The observed
number of collisions is proportional to the number of stars, implying
that our choice of scaling for time and stellar radius are
appropriate.  We therefore expect that the collision rate in our
models can be extrapolated to richer star clusters.  The star cluster
R\,136, for example, contains about 3 times as many stars as the
modeled clusters.  Scaling our results means that the collision rate in
R\,136 is about 8\, Myr$^{-1}$.

The absence of primordial binaries in our calculations possibly
affects the collision rate and the collision counterparts
significantly (see Kroupa 1997; 1998).\nocite{PK97}\nocite{PK98} It
is, however, not trivial to estimate the effect of the presence of a
large fraction of primordial binaries; apparently it is not even easy
to estimate the collision rate of a population of single stars
correctly.

\subsection{The blue stragglers in R\,136}
The small age of the star cluster R\,136 of $\sim 3$--4\,Myr and the
time needed for the first collisions to occur (about 1\,Myr) suggests
that about 20 collisions could have occurred in this cluster. The
result of these collisions should therefore still be visible, possibly
in the form of massive blue stragglers (Sandage 1953; Leonard \&
Duncan 1990; Leonard 1995; Sills et al.\
1997)\nocite{sand53}\nocite{1997ApJ...487..290S} in the cluster core.
\nocite{Lpjt95}\nocite{LpjtD90}

The three most massive stars in R\,136 have spectral type WN4.5 and
appear to be younger than the other stars.  Their age is estimated to
be about 1\,Myr (Massey \& Hunter 1998; de Koter et al.\
1997).\nocite{1997ApJ...477..792D} These stars show violet absorption
edges, which are common for late type (WN8 and later) stars but highly
unusual for these early types (Conti et al.\
1983).\nocite{1983ApJ...268..228C} Also striking is that these stars
are unusually hydrogen rich (Massey \& Hunter 1998).  They are about
an order of magnitude brighter than normal for such stars.  Estimates
for their masses range from 112 to 155\,\msun\ (Chlebowski \& Garmany
1991; Vacca et al.\ 1996).
\nocite{1991ApJ...368..241C}\nocite{1996ApJ...460..914V} Two of them
lie well inside the core of the cluster; the third is at a projected
distance of about 0.6\,pc from the core.

For a star cluster with an age of 4\,Myr these three massive stars
appear as blue stragglers.  It is therefore suggestive to identify the
three most massive stars in the star cluster R\,136 as collision
products.

\subsection{Black holes in dense star clusters}
When the runaway merger collapses to a black hole it is typically a
member of a rather close binary.  Upon dissociation of the binary, the
black hole is ejected from the core.  Since the compact object is
still considerably more massive than average, mass segregation brings
it back in the core within a few crossing times (see e.g.\,Hut,
McMillan, \& Romani 1992). \nocite{hmr92} 
New close binaries can be formed once the black
hole has returned to the core of the star cluster.  After an episode
of hardening the binary may become visible as an X-ray source when the
companion star starts to transfer mass to the black hole.  Such a
high-mass X-ray binary should be easily observable by X-ray
satellites. The age at which such a binary can form is at least $\sim
4$\,Myr, the minimum time needed for a black hole to form.  It is
likely to take considerably longer because the black hole has to
return to the core after its ejection.

The star cluster R\,136 is therefore ``too young'' for such a binary
to exist.  The star Mk\,34 at a distance of about 2.5\,pc from the
center of R\,136, however, is associated with a persistent X-ray
source with a luminosity of $\sim 10^{36}$\,erg\,s$^{-1}$ (Wang
1995).\nocite{1995ApJ...453..783W} Wang suggests that the binary
contains a black hole of between 2.4 and 15\,\msun\ accreting from the
dense wind of its spectral type WN4.5 Wolf-Rayet companion.  This
star, Mk\,34, can be classified as a ``blue straggler,'' as its
estimated age is about 1\,Myr, considerably smaller than the age of
the cluster (De Marchi et al.\ 1993).\nocite{DMNA+93}

Because R\,136 is too young for such an X-ray binary to be formed from
two collision products, it most likely formed from a primordial binary
ejected from the cluster core following the supernova which formed the
black hole.

\subsection{Collision rate}
The collision rates in our models are more than 10 times higher than
simple estimates based on cross sections.  In the computations with
12k stars, $14\pm4$ collisions occurred in a timespan of about 4\,Myr
whereas only $\sim1$ is expected.

Furthermore, the cross section arguments imply that low mass ($m\aplt
1$\,\msun) stars are most likely to collide.  In our simulations,
however, high-mass stars predominantly participated in encounters.
The most massive star participates in numerous collisions with other
stars.  Typically, the mass of this runaway grows to exceed
120\,\msun.  The rejuvenation of the runaway merger delays its
collapse to a compact object following a supernova.  Such a star could
be visible in the core of young star clusters with a high density as a
blue straggler.

The reason for the discrepancy between the formal cross-section
arguments and the results of our simulations is the neglect of mass
segregation and binary formation in the former estimates.  In the
simulations the most massive stars sink to the core due to dynamical
friction within a few half-mass crossing times, and form close
binaries by 3-body interactions.  The larger cross section of these
binaries increases the collision rate and makes them favored
candidates for encounters.

\acknowledgements We would like to thank Douglas Heggie and Atsushi
Kawai for discussions.  Edward P.J.\ van den Heuvel of the
Astronomical Institute ``Anton Pannekoek'' is acknowledged for
financial support.  SPZ thanks Drexel University for the hospitality
and Chris Colefax for his assistance.





\end{document}